\documentclass{iopart}

\usepackage[utf8]{inputenc}
\usepackage{url}
\usepackage{textcase}
\usepackage{bm}
\usepackage{graphicx}
\usepackage{textcomp}

\newcommand{\ket}[1]{\left|  #1  \right\rangle}
\newcommand{\bra}[1]{\left\langle  #1  \right|}

\newcommand{\expect}[1]{\left\langle #1 \right\rangle}

\newcommand{\K}{\, \mathrm{K}}

\newcommand{\T}{\, \mathrm{T}}

\newcommand{\rVec}[1]{\left( \!\! \begin{array}{cc} #1 \end{array} \!\! \right)}
\newcommand{\cVec}[1]{\left( \!\! \begin{array}{c} #1 \end{array}  \!\! \right)}
\newcommand{\myMat}[1]{\left( \!\! \begin{array}{cc} #1 \end{array}  \!\! \right)}

\begin{document}

\title{Model of the low temperature magnetic phases of gadolinium gallium garnet}

\author{M.~Ancliff$^1$, V.~Sachnev$^2$ and N.~d'Ambrumenil$^3$}
\address{$^1$ Department of Physics, The Catholic University of Korea, Bucheon, Gyeonggi-Do, South Korea}
\address{$^2$ Department of Information, Communication, and Electronic Engineering, The Catholic University of Korea, Bucheon, Gyeonggi-Do, South Korea}
\address{$^3$ Department of Physics, University of Warwick, Coventry CV4 7AL, United Kingdom}

\begin{abstract}
    
    The magnetic behaviour of gadolinium gallium garnet in an external magnetic field at zero temperature is considered. For high fields a classical spin model of the gadolinium ions predicts a spin configuration that is periodic at the level of the smallest repeating unit cell. The quantum version of the model is treated via a truncated Holstein-Primakoff transformation with axes defined by the classical spin configuration, and the magnon excitation bands are calculated. The model predicts a transition in the field range of $1.9$---$2.1\T$, sensitive to the direction of the applied field, which is caused by one or more magnon modes becoming soft as the field is decreased. In general the soft modes occur at incommensurate wavevectors and therefore break the periodicity of the spin configuration below the transition. One exception occurs when the field aligns with one of the principle crystal axes, in which case periodicity of the spin configuration is found to be maintained on a larger crystallographic cubic cell even below the transition. This simple case is studied in more detail. Comparisons are drawn with existing experimental data, and further experimental tests of the model are suggested.
    
\end{abstract}

\vspace{5cm}
\begin{quote}
This is the version of the article before peer review or editing, as submitted by an author to Journal of Physics: Condensed Matter. IOP Publishing Ltd is not responsible for any errors or omissions in this version of the manuscript or any version derived from it. The Version of Record is available online at \underline{https://doi.org/10.1088/1361-648X/ac1f4f}.
\end{quote}

\maketitle

\section{Introduction}

Gadolinium gallium garnet ($\mathrm{Gd}_3\mathrm{Ga}_5\mathrm{O}_{12}$) has attracted interest as a frustrated anti-ferromagnetic crystal displaying a large number of different magnetic phases as a function of temperature and applied magnetic field \cite{Hov1980,Ramirez1991,Tsui1999,Deen2015,Rousseau2017}. At low external fields and intermediate temperatures the system displays a spin-liquid phase with hidden long-range order found in the spin configurations around 10-site rings of gadolinium ions \cite{Paddison2015}. The system undergoes an unconventional spin-glass transition as the temperature is lowered below around $0.15\K$ \cite{Schiffer1995, Petrenko1998,Quilliam2013}, whose nature is not completely understood, but which has been suggested to be related to a change in the ordering around the same 10-site rings \cite{Ghosh2008}. At large fields (around $1.8\T$ and above) there is a paramagnetic phase in which spins are closely aligned with the magnetic field, with some periodic canting due to the spin-spin interactions \cite{dAmbrumenil2015}. 

The behaviour of the system between the spin-liquid and paramagnetic phases has been probed by measurements of magnetic susceptibility\cite{Hov1980,Schiffer1994}, specific heat and thermal expansion\cite{Ramirez1991,Tsui1999}, neutron diffraction\cite{Deen2015,Petrenko2002,Petrenko1999,Petrenko2009}, and sound velocity and attenuation\cite{Rousseau2017}. These experiments reveal several intermediate field-induced antiferromagnetic phases, some of which display spin ordering incommensurate with the lattice, but which cannot be explained in terms of a single spin-wave \cite{dAmbrumenil2015,Petrenko2009}. Much is still unknown about the magnetic ordering in these intermediate phases, but evidence shows they are sensitive to the direction of the applied magnetic field relative to the crystal axes \cite{Hov1980}. This strong anisotropy of the phase diagram has been most clearly shown in comparisons of the cases with $B||(001)$ and $B||(110)$, which can be found in \cite{Rousseau2017}, \cite{Petrenko2002}, and \cite{Petrenko2009}.

This paper presents a theoretical study of the transition between the high-field
paramagnetic phase to the intermediate phase below it at zero temperature. The long-range dipolar forces are found to play a critical role in the nature of the transition and this is suggested as an explanation for the anisotropy seen in the phase diagram. The theoretical model is introduced in section \ref{sec:theoreticalModel}. The analysis in the high field  regime is simplified as the spin configurations are known to be periodic at the level of the smallest repeating unit cell of the garnet lattice. In section \ref{sec:high-field} the nature of the transition into the intermediate phase at lower applied fields is studied, and is found to be caused by the softening of one or more magnon modes in the paramagnetic phase. The critical field and the wavevector of the soft modes depend on the field direction and, in the case of $B||(110)$, a whole line of modes is found to go soft at approximately the same field. For most field directions the soft modes occur at wavevectors incommensurate with the reciprocal lattice, which breaks the periodicity and suggests the creation of incommensurate spinwaves below the transition. The case of $B||(001)$ is unusual in that a single magnon mode goes soft at the H-point of the first Brillouin zone, suggesting that the spin ordering may remain commensurate with the lattice below the transition, which allows a more detailed examination of this case in section \ref{sec:Bz}. The results of the model are compared with the known experimental results, and suggestions are made for neutron scattering measurements which could further test the predictions of the model.

\section{Theoretical model and methods used} \label{sec:theoreticalModel}

The spin-states of the $\mathrm{Ga}^{2+}$ and $\mathrm{Gd}^{3+}$ ions in gadolinium gallium garnet are $S=L=0$ and $S=7/2, L=0$ respectively. Thus, magnetic properties of gadolinium gallium garnet are driven by the $\mathrm{Gd}^{3+}$ ions, which display significant antiferromagnetic exchange and long-range dipolar interactions. The gadolinium ions lie on a garnet lattice, belonging to the BCC crystal class, with 12 gadolinium ions per smallest repeating unit cell, henceforth referred to as the BCC unit cell. The smallest repeating \emph{cubic} cell of the lattice, with double the volume of the BCC unit cell, has side length $a_0 = 12.35$\AA \, \cite{Petrenko1998}, and in what follows is referred to as the unit cubic cell.

\begin{figure}
    \begin{indented}
    \item[]
    \includegraphics[width=8cm]{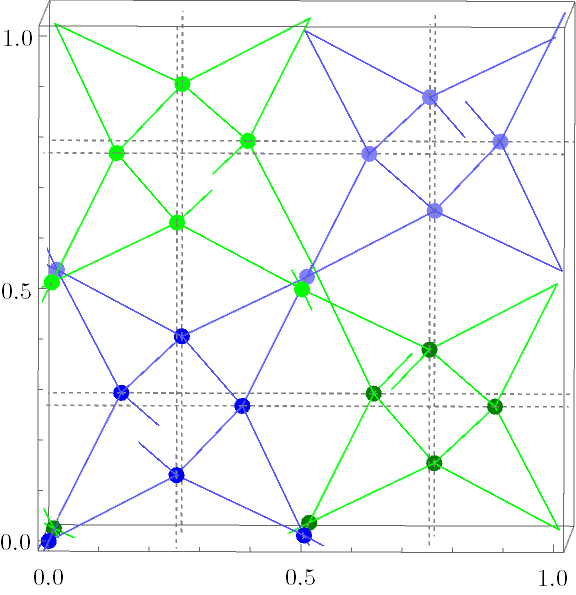}
    \caption{Positions of the 24 gadolinium ions in one crystallographic unit cubic cell. The nearest neighbour graph splits into two disconnected sublattices, here shown in blue and green. Third-nearest neighbour interactions are shown as dashed gray lines. Spins on the two sublattices are connected only by third nearest neighbour and dipolar interactions.}
    \label{fig:GGGstructure}
    \end{indented}
\end{figure}

The nearest neighbour graph of the gadolinium ions shows two unconnected sublattices composed of equilateral triangles connected at the vertices. The nearest neighbour antiferromagnetic interactions therefore give rise to frustration. The two sublattices are coupled by interactions of third-nearest neighbours and the dipole-dipole interaction. See figure \ref{fig:GGGstructure} for a representation of the lattice structure and interactions.

Adiabatic susceptibility measurements \cite{Kinney1979} suggest an effective Hamiltonian with Zeeman, antiferromagnetic, and dipolar terms,
\begin{eqnarray} 
 H & = & -g \mu_B \mathbf{B} \cdot \sum_j \mathbf{S}_j + \frac{1}{2} \sum_{i,j \atop i\neq j} J(|\mathbf{r}_{ij}|) \mathbf{S}_i \cdot \mathbf{S}_j + \nonumber \\
  & &  + \frac{D}{2} \sum_{i,j \atop  i \neq j} \left(
    \frac{\mathbf{S}_i \cdot \mathbf{S}_j}{|\mathbf{r}_{ij}|^3} - 3 \frac{(\mathbf{S}_i \cdot \mathbf{r}_{ij})
    (\mathbf{S}_j \cdot \mathbf{r}_{ij})}{|\mathbf{r}_{ij}|^5} \right) \mathrm{,} 
    \label{eq:hamiltonian}
\end{eqnarray}
where $J(|\mathbf{r}_{ij}|)$ determines the strength of spin-spin interactions, and only interactions between first-, second-, and third- nearest neighbors are assumed to be significant. The model assumes the spins are isotropic (the large, pure spin state of the gadolinium ions should minimize the effect of the crystal field). Experimental measurements have provided an upper-bound on the single-ion spin anisotropy of $0.04\K$ \cite{Overmeyer1963}.

The value of the dipolar interaction strength is $D {r_{nn}}^{-3} = 0.046\K$ where $r_{nn}$ denotes the nearest-neighbour distance. Best-fit values for the coupling constants were found to be $J_1 = 0.107 \K$, $J_2 = - 0.003 \K$ and $J_3 = 0.013 \K$, though the study found significant uncertainty in the values of $J_1$ and $J_2$ as a function of $J_3$ \cite{Kinney1979}. A later study refined the values of $J_2 = - 0.005\K$ and $J_3 = 0.010 \K$ using neutron scattering data and assuming the same value for $J_1$ \cite{Yavorskii2006}. The difference in the quoted values of $J_2$ and $J_3$ have minimal impact on the findings presented in this paper.

Models with the Hamiltonian (\ref{eq:hamiltonian}) have been employed in Monte-Carlo simulations \cite{Paddison2015,Petrenko1999,Schiffer1994,Petrenko2000} and mean-field treatments \cite{Yavorskii2006,Yavorskii2007} of the system, which have been used to probe both the low-field spin-liquid and high-field paramagnetic phases of the system. In the high-field phase the model has been shown to be capable of explaining puzzling experimental findings such as the appearance of non-ferromagnetic Bragg peaks in neutron scattering experiments \cite{dAmbrumenil2015}. However, the Monte-Carlo simulations cited above relied on short-range cutoffs of the dipolar term (at third- or fourth- nearest neighbours) which has the potential to lead to significant errors, especially given the importance of the dipolar term in determining the spin-ordering below the transition. (See also the sensitivity of simulations of the zero-field phase to finite range cutoffs reported in \cite{Yavorskii2007}.) In this paper the dipolar term is evaluated using the Ewald method \cite{Ewald1921,deLeeuw1980, Wang2001} to avoid these issues. 

Before describing the full model it is useful to discuss the simplified case with no dipolar or spin-spin interactions beyond the nearest neighbour. Here the model displays complete rotational symmetry (a simultaneous rotation of all spins and Zeeman term commutes with the Hamiltonian), so the direction of the crystal axes with respect to that of the magnetic field is unimportant, and the fully ferromagnetic state with all spins aligned with the external field is an eigenstate of the Hamiltonian. The magnon spectrum from this eigenstate shows that the lowest-energy excitations form a 4-fold degenerate flat band, due to the existence of perfectly localized excitations around the 10-site rings mentioned above \cite{dAmbrumenil2015,Bergman2008}. The ferromagnetic eigenstate is found to be stable to magnon-excitations down to a field of around $1.7\T$, where the degenerate flat bands go soft. Including either dipolar or further spin-spin interactions into the model destroys the exact localization of the excitations and distorts the bands. The dipolar term also breaks the rotational symmetry and means that the ferromagnetic state is no longer an exact eigenstate of the Hamiltonian.

The large spin of the $\mathrm{Gd}^{3+}$ ions suggests that a classical spin model is a reasonable starting point. We take spins of length $7/2$, and use the Hamiltonian (\ref{eq:hamiltonian}) with the values from \cite{Kinney1979}. It is assumed that the spin configuration in the ground state of the high-field phase has the same periodicity as the lattice. This assumption is supported by neutron scattering data for integer peaks such as $(111)$ and $(210)$, which are ruled out for configurations which are periodic on the level of the BCC unit cell. Experimentally, such peaks are absent in the high-field phase but appear below the transition \cite{Deen2015,dAmbrumenil2015}. To further test this assumption we performed numerical optimization on systems up to $30 \times 30 \times 30$ unit cubic cells in the high-field phase. In each case a unique minimal energy state was found in which each BCC unit cell had identical spin configuration, as assumed.

Above the transition, an approximate ground-state for the quantum system and corresponding magnon excitation spectrum are calculated following a similar method to that presented in \cite{DelMaestro2004}. First the classical ground-state spin configuration for a single BCC unit cell with periodic boundary conditions is calculated numerically. Local axes $(\mathbf{e}^x_i,\mathbf{e}^y_i,\mathbf{e}^z_i)$, are chosen at each site (i=1,\ldots,12) of the BCC unit cell, so that the local $z$-axis is aligned with the classical ground-state spin at the same site. A Holstein-Primakoff transformation based upon these axes is then applied to the quantum Hamiltonian, truncated at zeroth order in $S$,
\begin{eqnarray}
    S^x_{A,i} & \to & (S/2)^{1/2} (\hat{a}_{A,i} + {\hat{a}_{A,i}}^\dagger) \nonumber \\
    S^y_{A,i} & \to & -\rmi (S/2)^{1/2} (\hat{a}_{A,i} - {\hat{a}_{A,i}}^\dagger) \label{eq:hpTrans} \\
    S^z_{A,i} & \to & S - {\hat{a}_{A,i}}^\dagger \hat{a}_{A,i} \nonumber
\end{eqnarray}
where $S=7/2$ and $A$ indexes the BCC unit cell in the lattice. Fourier transforming the cell index, $a_{A,i} \to a_i(\mathbf{q})$ leads to a Hamiltonian of the form,
\begin{equation}\label{eq:hamHolsPrima}
    H \! = \! \sum_\mathbf{q} 
    \rVec{ a_i (\mathbf{q}) \!\! &  \!\! {a_i}^\dagger (\mathbf{q})}
    \!\! \myMat{h^\mathrm{ac}_{ij}(\mathbf{q}) \! & \! h^\mathrm{aa}_{ij}(\mathbf{q}) \\
    h^\mathrm{cc}_{ij}(\mathbf{q}) \! & \! h^\mathrm{ca}_{ij}(\mathbf{q})}
    \!\! \cVec{{a_j}^\dagger (\mathbf{q}) \\ {a_j} (\mathbf{q})} 
\end{equation}
where summation over the site indices $i,j$ is implied. The dipolar and exchange interactions give rise to particle-number non-conserving terms, $({\hat{a}_i}^\dagger {\hat{a}_j}^\dagger + \mathrm{h.c.})$ (such terms arise in the exchange interaction due to the choice of non-parallel local axes). These terms can be removed, and the Hamiltonian diagonalized via a Bogoliubov transformation,
\begin{equation}
    \hat{b}_j(\mathbf{q}) = \left( c_j^i \hat{a}_i(\mathbf{q}) + \bar{c}_j^i {\hat{a}_i}^\dagger(\mathbf{q}) \right)
\end{equation}
for some coefficients $c_j^i$ and $\bar{c}_j^i$ (note $\bar{c}_j^i$ is not the complex conjugate of $c_j^i$). The resulting Hamiltonian can be written
\begin{equation}
    H = \sum_\mathbf{q} \sum_{\alpha=1}^{12} \epsilon_\alpha(\mathbf{q}) \left( {\hat{b}_\alpha}^\dagger (\mathbf{q}) \hat{b}_\alpha (\mathbf{q}) + \frac{1}{2} \right) \mathrm{.}
\end{equation}
The quantum ground state, $\ket{\tilde{0}}$, is taken to be the state annihilated by all $\hat{b}_i(\mathbf{q})$. The magnon spectrum is then given by the function $\epsilon_i(\mathbf{q})$. 

The Ewald technique must be adapted to handle spinwaves with non-zero $\mathbf{q}$ to calculate the correct matrix elements in equation (\ref{eq:hamHolsPrima}). This results in a non-analyticity in the matrix elements when $\mathbf{q}$ is equal to a reciprocal lattice vector, related to the fact that in an infinite lattice the sum of the dipolar terms is not absolutely convergent. A method by which the sum can be regularized is described in \cite{deLeeuw1980}. As a consequence,
some magnon bands become discontinuous at the origin in the thermodynamic limit
(see the highest energy bands in figure \ref{fig:magnonBands} for an example of this). The mathematical details regarding the magnon spectrum are discussed in Appendix C of \cite{Gingras2004} (where the non-analyticity is described by equations (C23) and (C25) therein). While this is an
interesting phenomenon, by inspection the lowest lying bands do not develop any such discontinuities, and so this effect has no impact on the transition.

From the quantum ground state and magnon spectrum simulated values for elastic and inelastic neutron scattering intensities can be found \cite{Jensen1991Book}. The elastic (Bragg) scattering intensity at wavevector $\mathbf{k}$ is calculated as
\begin{equation}
  \fl  |F(\mathrm{k})|^2  \sum_{i,j} 
   \left( \mathbf{e}_i^z \cdot \mathbf{e}_j^z - (\mathbf{k} \cdot \mathbf{e}_i^z ) (\mathbf{k} \cdot \mathbf{e}_j^z ) / |\mathrm{k}|^2 \right)  \sum_{A,B} \rme^{-\rmi \mathbf{k} \cdot (\mathbf{r}_{A,i} - \mathbf{r}_{B,j})}\expect{S_{A,i}^z} \expect{S_{B,j}^z} \mathrm{,}
   \label{eq:Elastic}
\end{equation}
where $F(\mathrm{k})$ is the magnetic form factor of the $\mathrm{Gd}^{3+}$ ions, $\langle \hat{O} \rangle = \bra{\tilde{0}} \hat{O} \ket{\tilde{0}}$ denotes the ground state expectation of operator $\hat{O}$, and
$S_{A,i}^\mu$ is as in equation \ref{eq:hpTrans}. The inelastic intensity at wavevector $\mathbf{k}$ and energy $\epsilon$ is calculated as
\begin{eqnarray}
   \fl |F(\mathrm{k})|^2 \sum_{\mu,\nu} \sum_{i,j}  
    \left( \mathbf{e}_i^\mu \cdot \mathbf{e}_j^\nu - (\mathbf{k} \cdot \mathbf{e}_i^\mu ) (\mathbf{k} \cdot \mathbf{e}_j^\nu ) / |\mathrm{k}|^2 \right) \sum_{A,B} \rme^{-\rmi \mathbf{k} \cdot (\mathbf{r}_{A,i} - \mathbf{r}_{B,j})} \nonumber \\
     \quad \times
     \sum_\mathbf{q} \sum_{\alpha=1}^{12} 
     \expect{S_{A,i}^\mu {\hat{b}_\alpha}^\dagger (\mathbf{q})}
     \expect{ {\hat{b}_\alpha} (\mathbf{q}) S_{b,j}^\nu }
     \delta (\epsilon - \epsilon_\alpha (\mathbf{q})) \mathrm{,}
     \label{eq:inelastic} 
\end{eqnarray}
where $\mu, \nu \in \{ x,y \}$.

\section{The high-field phase} \label{sec:high-field}

\begin{figure}
    \begin{indented}
    \item[]
    \includegraphics[width=7cm]{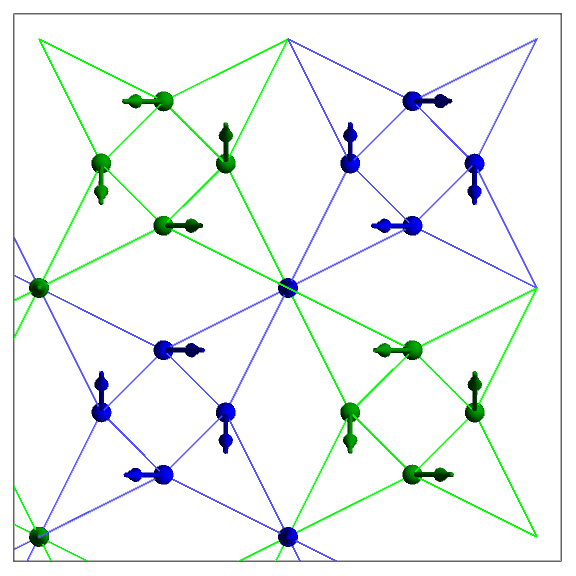}
    \caption{Canting of the classical spins for $B||(001)$ at a field of $2.0\T$ (just above the transition). The viewpoint is looking down along the $z$-axis. The spin configurations above the transition share the $D_4$ point symmetry group of the Hamiltonian, and as a consequence the spins closest to the edges and at the centre of the cell shown are exactly parallel to the field (four additional sites are hidden behind these sites in the projection shown here -- compare with figure \ref{fig:GGGstructure}).}.
    \label{fig:spinCanting}
    \end{indented}
\end{figure}

In the high field phase the classical spins remain closely aligned with the external field, but there is some small angle canting due to the dipolar and antiferromagnetic interactions. Figure \ref{fig:spinCanting} shows an example for $B||(001)$ just above the transition.

\begin{figure*}
    \begin{indented}
    \item[]
    \begin{tabular}{ll}
    (a) & (c) \\
    \includegraphics[width=4.5cm]{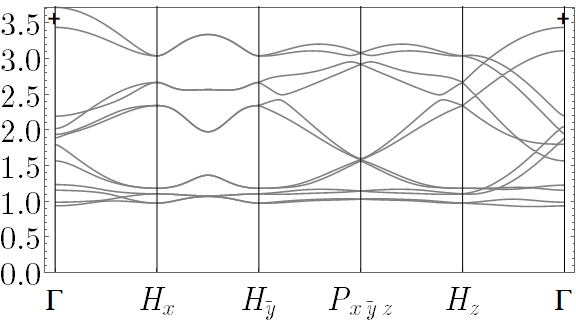} &
    \includegraphics[width=4.5cm]{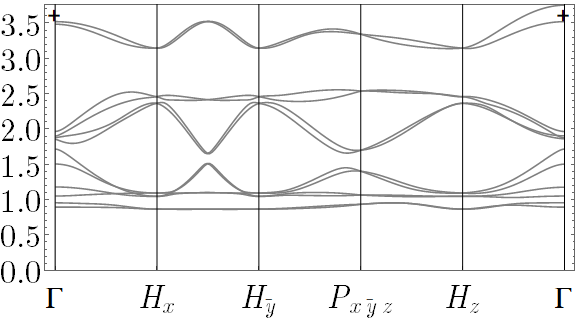} \\
    (b) & (d) \\
    \includegraphics[width=4.5cm]{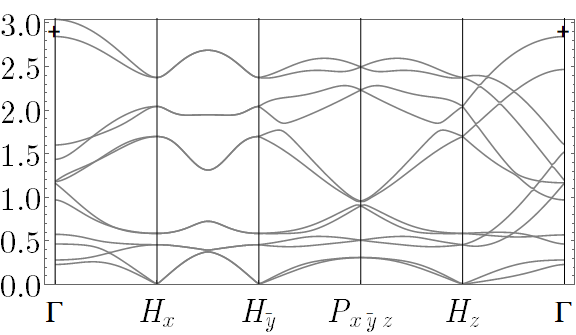} &
    \includegraphics[width=4.5cm]{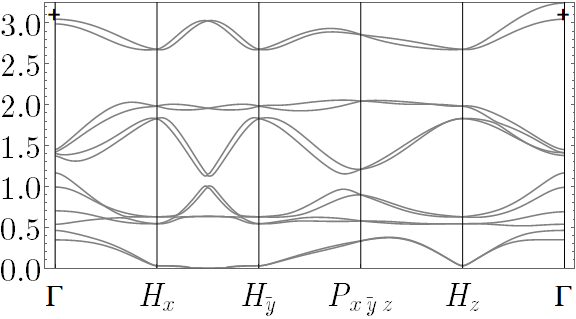}
    \end{tabular}
    \caption{Magnon excitation bands for: (a) $B||(001)$ at $2.5\T$, (b) $B||(001)$ at the transition, (c) $B||(110)$ at $2.5\T$, (d) $B||(110)$ at the transition. The vertical axis shows excitation energy in Kelvin. For both field directions the lowest band is approximately dispersionless at a field of $2.5\T$ but develops significant dispersion at the transition. In units of the reciprocal cubic lattice length, $2\pi /a_0$, the labeled points are $H_x=(1,0,0)$, $H_{\overline{y}}=(0,-1,0)$, $P_{x \overline{y} z} = (1,-1,1)/2$, and $H_z=(0,0,1)$ (all $H$ points are equivalent but the directions of approach to them are not). The discontinuity in the energies of some bands as $q$ approaches the $\Gamma$ point is real -- see the text for a discussion. The excitation energies close to, and at, $\Gamma$ and reciprocal lattice vectors are, however, all well-defined in the thermodynamic limit. In the highest energy band, where the discontinuity is most pronounced, the energy at the $\Gamma$ point is indicated by a cross.}
    \label{fig:magnonBands}
    \end{indented}
\end{figure*}

\begin{figure}
    \begin{indented}
    \item[]
    \includegraphics[width=7cm]{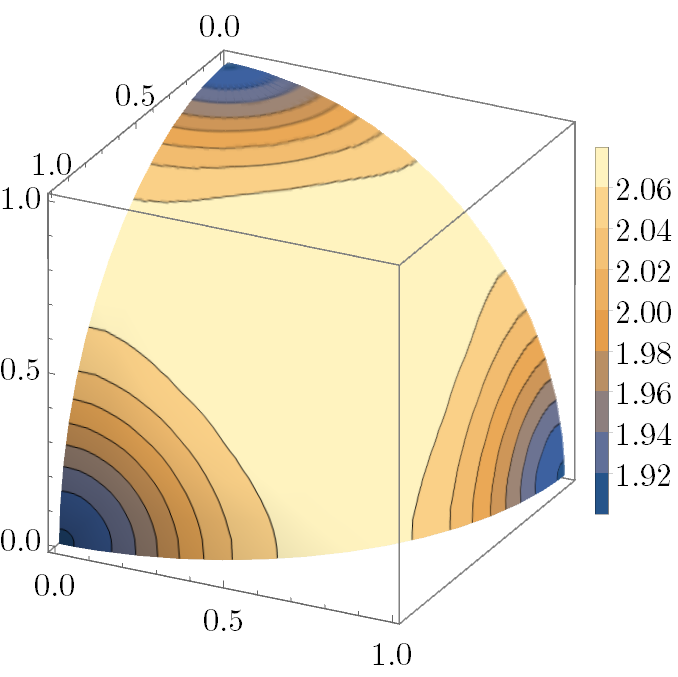}
    \caption{Dependence of the critical field strength on field direction relative to the crystal axes. Field directions are represented as points on the unit sphere, and the contours show the critical field at each point.}
    \label{fig:criticalField}
    \end{indented}
\end{figure}

At fields around $2.5\T$ and above, the lowest magnon band is approximately flat, but the distortion increases significantly as the transition is approached, and at the transition the band goes soft at a particular wavevector $\mathbf{q}$, although for some field directions several modes go soft at approximately the same field. See figure \ref{fig:magnonBands} for examples when $B||(001)$ and $B||(110)$. The magnitude of the critical field depends upon its orientation, as shown in figure \ref{fig:criticalField}, and is minimum when the field aligns with a crystal axis and maximum when it lies along the $(111)$ direction. Experiments have also found a directional dependence of the critical field, but the field predicted by the model is around 25\% above the values found by single-crystal neutron scattering data. Possible reasons for this discrepancy will be discussed in section \ref{sec:discuss}. Specifically, for $B||(0,0,1)$ predicted and experimentally observed critical fields are $B_\mathrm{pred} = 1.92 \T$ and $B_\mathrm{exp} = 1.5\T$ \cite{Petrenko2002}, respectively., whereas for  $B||(1,1,0)$ the corresponding values are $B_\mathrm{pred} = 2.07 \T$ and $B_\mathrm{exp} = 1.67 \T$ \cite{Petrenko2009}. The ratios $B_\mathrm{exp}/B_\mathrm{pred}$ are approximately the same in both cases.

For most directions of the magnetic field the magnon band is found to go soft at an incommensurate wavevector in the vicinity of the $H$-point of the first Brillouin zone. This implies that the ground state spin configurations of different BCC unit cells will no longer be the same below the transition. In other words the transition is characterized by the appearance of soft magnon modes which break the BCC-periodicity of the spins. Cases $B||(001)$ and $B||(110)$, which have been most thoroughly studied in single-crystal experiments, are unusual in terms of the position of the soft modes. For $B||(001)$ the band is found to go soft exactly at the $H$-point, whereas for $B||(110)$ a whole line of modes between two $H$-points go soft at approximately the same field, see figure \ref{fig:magnonBands}.

The prediction of a second-order transition driven by the softening of magnon modes is consistent with the sharp peaks seen in magnetic susceptibility and specific heat measurements of the transition \cite{Ramirez1991,Deen2015,Schiffer1994}. Magneto-calorific measurements also show no measurable latent heat associated with the transition (whereas latent heats are observed in the further transitions at lower fields) \cite{Tsui1999}. 

\begin{figure*}
    \begin{indented}
    \item[]
    \begin{tabular}{ll}
    (a) & (b) \\
    \includegraphics[width=5cm]{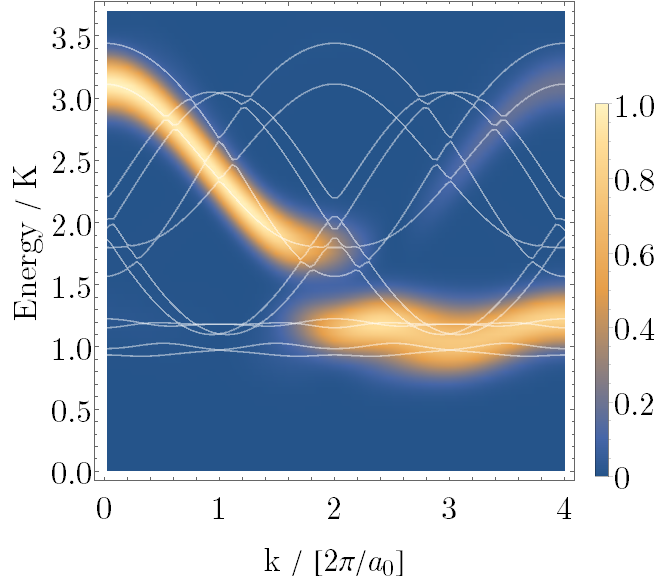} &
    \includegraphics[width=5cm]{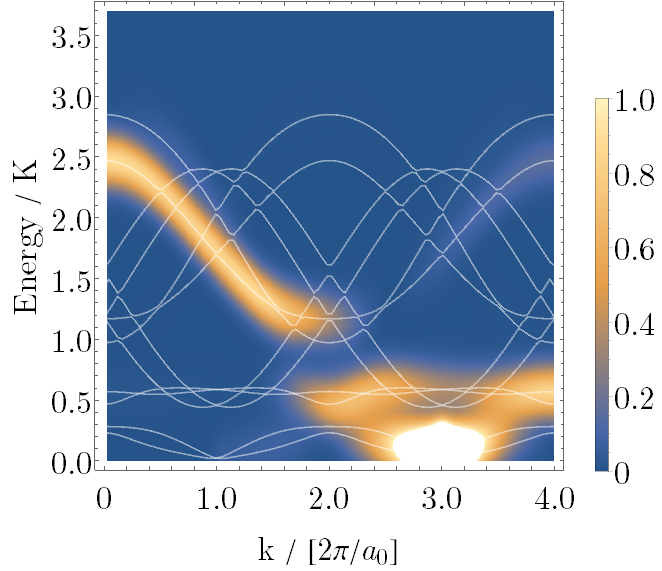}
    \end{tabular}
    \caption{Simulated inelastic neutron scattering intensities for $B||(001)$ at fields of (a) $2.5\T$ and (b) just above the transition. The scattering wavevector $\mathbf{k}$ runs from the origin to the point $(0,0,4)$ in units of $2\pi / a_0$. The positions of the magnon bands are overlayed in white. Strong quasi-elastic scattering is caused by the nearly soft mode at $(0,0,3)$ in the second diagram. (The predicted scattering intensity at this point increases by a factor of ten as the transition is approached, as a consequence of an increase in magnitude of the expectations $\expect{ {\hat{b}_\alpha} (\mathbf{q}) S_{b,j}^\nu }$ occuring in equation (\ref{eq:inelastic}) for the lowest two magnon bands.)}
    \label{fig:inelasticScatteringBz}
    \end{indented}
\end{figure*}

\begin{figure}
    \begin{indented}
    \item[]
    \includegraphics[width=6cm]{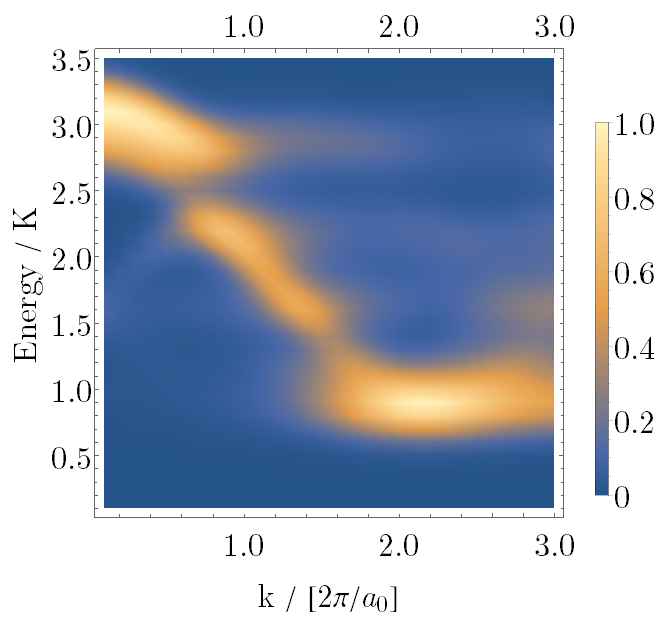}
    \caption{Inelastic scattering intensity at an external field of $2.5\T$. Directions of magnetic field and momentum vectors have been averaged over to allow direct comparison to existing experimental data from a powder sample (see \cite{dAmbrumenil2015} fig. 2).}
    \label{fig:inelasticScatteringPowder}
    \end{indented}
\end{figure}

Simulated inelastic neutron scattering intensities suggest that the appearance of the soft mode should be clearly visible as the transition is approached -- see figure \ref{fig:inelasticScatteringBz} for the case of $B||(001)$ -- which gives another way to test the key prediction of our analysis. We have also calculated simulated inelastic scattering intensities with an average over magnetic field and wavevector directions, to allow comparison with existing data from a powder sample \cite{dAmbrumenil2015}. The result is shown in figure \ref{fig:inelasticScatteringPowder} and shows reasonable agreement with the experimental data.

\begin{figure}
    \begin{center}
    \begin{tabular}{lll}
    (a) & (b) & (c)  \\
    \includegraphics[width=4cm]{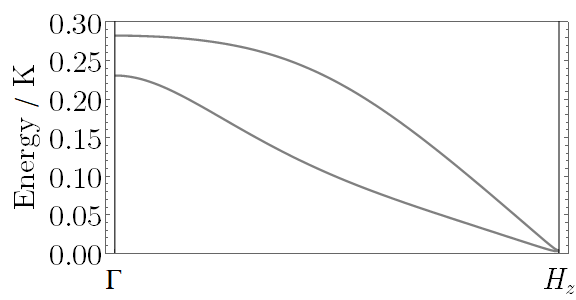} &
    \includegraphics[width=4cm]{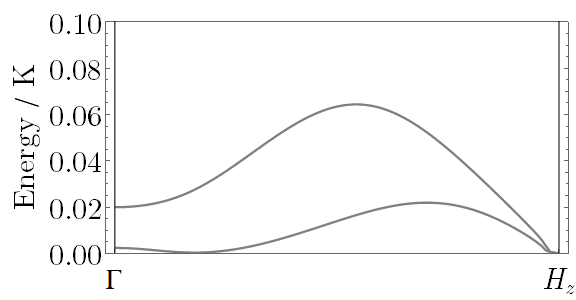} &
    \includegraphics[width=4cm]{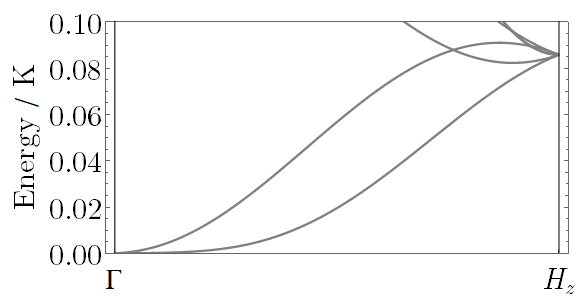} 
    \end{tabular}
    \end{center}

    \caption{Expanded view of the dispersion of the lowest magnon bands at the transition field for three different values of $D {r_{nn}}^{-3}$: (a) $D {r_{nn}}^{-3} = 0.046\K$ as stated in the text, (b) $D {r_{nn}}^{-3} = 0.016\K$, and (c) $D {r_{nn}}^{-3} = 0$ (no  dipolar interaction). The position of the soft magnon mode moves discontinuously from the H point to a position near the $\Gamma$ point as $D {r_{nn}}^{-3}$ decreases through the value $0.016$ (panel b).}
    \label{fig:magnonsDifferentD}
\end{figure}

The position of the soft mode is found to be sensitive to the $D/J_1$ ratio as well as the field direction. If the value of the dipolar coupling, $D$, is decreased to around a third of the estimate given in section \ref{sec:theoreticalModel}, the position of the soft mode changes discontinuously -- from a position near the $H$ point to a position near the $\Gamma$ point, as shown in figure \ref{fig:magnonsDifferentD}. This implies that approaches ignoring the dipolar interaction or treating it as a perturbation are unlikely to capture the correct nature of the phase(s) below the transition. Other gadolinium garnets, such as gadolinium aluminium garnet and gadolinium tellurium lithium garnet have the same gadolinium lattice structure but different $D/J_1$ ratios, and so provide a way to test the dependence of the transition on this ratio \cite{Quilliam2013}.

\section{Below the transition} \label{sec:Bz}

The finding that the transition is driven by a soft magnon mode which breaks the periodicity of the spins is consistent with the appearance of Bragg scattering peaks at incommensurate wavevectors seen in neutron scattering experiments below the transition. For $B||(110)$ the soft wavevectors are predicted to be of the form $(1-a,a,0)$ for $0<a<1$, and while single-crystal neutron scattering data is only available for the $(kkl)$ plane, a peak at $(\frac{1}{2},\frac{1}{2},2)$, consistent with the development of a spinwave at $\mathbf{k}=(\frac{1}{2},\frac{1}{2},0)$, has been observed \cite{Petrenko2009}. It should be noted that the same experiment also observed incommensurate peaks at $(1.18,1.18,1.32)$, $(-1.53,-1.53,0.51)$, $(0.59,0.59,1.95)$, and $(1.46,1.46,-0.51)$, which do not correspond to soft-mode wavevectors predicted at the transition. The presence of these additional peaks and the theoretical prediction of many spinwave excitations with very similar low energies at the transition suggest that magnon-magnon interactions are important for understanding the intermediate phase. These interactions have been ignored in our analysis, but are of order $J_1 |S|^0 \sim 0.1 \K$, which is comparable to the energy-width of the lowest band shown in figure \ref{fig:magnonBands}(d).

If the ground-state classical spin configuration no longer shares the same periodicity as the lattice, the treatment outlined in section \ref{sec:theoreticalModel} will no longer work. Looking for approximate periodic solutions with periods larger than a few unit cells becomes impractical given the computational complexities in the calculation of the approximate quantum ground state and magnon modes. In contrast to the situation in the high-field phase, numerical optimization of the classical spin model below the transition reveals
a large number of local energy minima, which is also an obstacle to the study of the lower-field phases. Thus, a thorough treatment of the intermediate phase for general field direction is a harder problem, and is not attempted here.

One exception is the case when the external field aligns with one of the crystal
axes. In this case a single soft mode appears exactly at the $H$ point of the first Brillouin zone, and below the transition the spins may therefore remain periodic on the larger unit cubic cell. In this case the methods of section \ref{sec:theoreticalModel} can be applied to this larger cell. In what follows we explore the predictions of the model for $B||(001)$ below the transition using this assumption.

\begin{figure}
    \begin{center}
    \includegraphics[width=14cm]{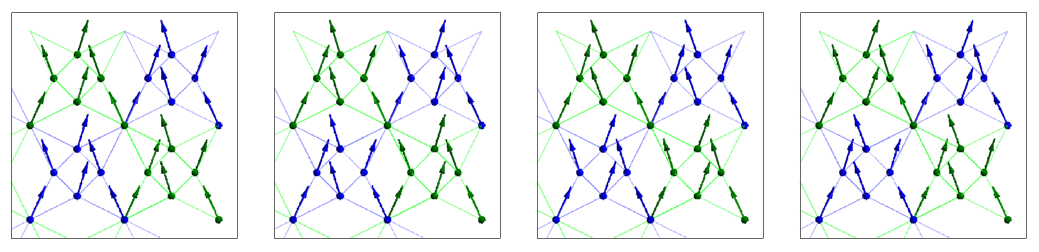} 
    \end{center}
    \caption{The spin configurations of the four minimum energy solutions at a field of $1.7\T$. The first two configurations are equivalent up to translation, as are the last two. The first pair of solutions can be mapped onto the second by a $90^\circ$ rotation around and translation along the $z$-axis. The viewpoint is along the $x$-axis with the $z$-axis vertical.}
    \label{fig:spinsBz}
    
\end{figure}

For fields between the transition (at 1.92T in our approach) and 0.3T, there are four degenerate classical spin configurations that minimise the energy. Similar solutions have been seen before in Monte-Carlo simulations of the same problem \cite{Schiffer1994}, and are illustrated in figure \ref{fig:spinsBz}. Two pairs of these are related by a translation by $\left( \frac{1}{2} , \frac{1}{2}, \frac{1}{2} \right)$, so modulo translations there are only two distinct solutions. These solutions can be mapped into each other by a $90^\circ$ rotation around the $z$-axis followed by a translation parallel to the same axis. 

\begin{figure}
    \begin{indented}
    \item[]
    \includegraphics[width=4.5cm]{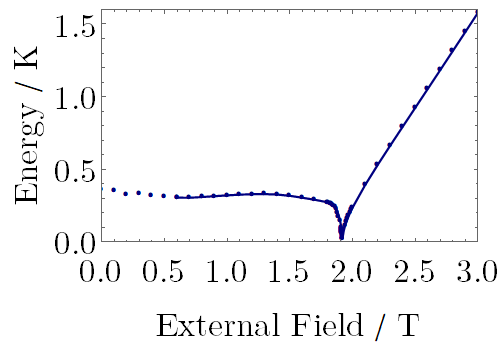}
    \caption{Minimum magnon excitation energy as a function of external field strength for $B||(001)$. The excitation gap falls to zero as the field approaches the transition value of $1.92 \T$, both from above and from below. Away from the transition at $1.92\T$ the periodic spin configuration on the unit cubic cell is found to be stable to spinwave excitations, with the energy gap remaining around $0.3\K$. Below around $0.5\K$, while the spinwave gap remains roughly constant, there exist non-spinwave excitations with lower excitation energy, see the text for more details.}
    \label{fig:minimumMagnon}
    \end{indented}
\end{figure}

Calculations of the magnon spectra show that the magnon excitation energy gap is non-zero except at the transition. This means away from the transition the solution is stable to spinwave excitations and therefore the assumption about the periodicity of the spins below the transition is internally consistent in this regard. The energy of the lowest magnon mode as a function of field strength is shown in figure $\ref{fig:minimumMagnon}$. To further test the assumption of periodicity we performed numerical optimization of the classical spins on lattice sizes up to $20\times 20 \times 20$ unit cubic cells. While a large number of locally optimal configurations were found, the periodic configurations were found to have the minimum energy of all local optima discovered.

\begin{figure}
    \begin{center}
    \begin{tabular}{lll}
    (a) & (b) & (c) \\
    \includegraphics[width=3.8cm]{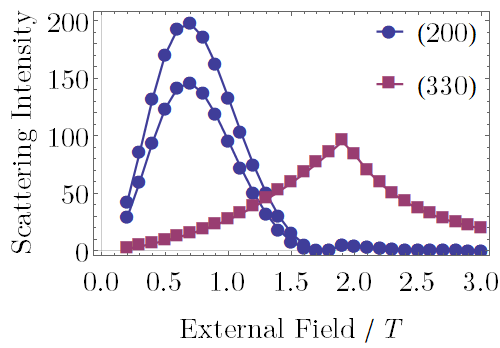} &
    \includegraphics[width=3.8cm]{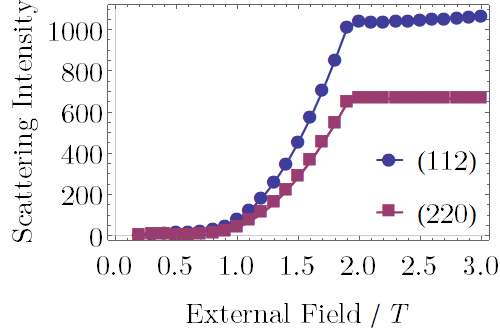} &
    \includegraphics[width=3.8cm]{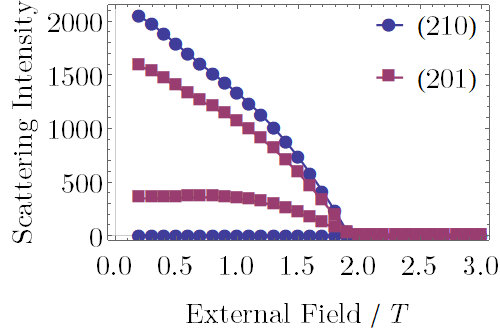} 
    \end{tabular}
    \end{center}
    \caption{Predicted neutron scattering profiles for: (a) the antiferromagnetic peaks $(200)$ and $(330)$, (b) the ferromagnetic peaks $(112)$ and $(220)$, and (c) the antiferromagnetic peaks $(210)$ and $(201)$. Scattering intensity units are arbitrary but can be compared between graphs. The peaks $(200)$, $(210)$ and $(201)$ show two different profiles corresponding to the two possible spin configurations below the transition, see the text for more details. Experimental data is available for the peaks in (a) and (b) \cite{Petrenko2002}.}
    \label{fig:elasticScatterBz}
\end{figure}

Above the transition the spin configuration is symmetric under $90^\circ$ rotations around the $z$-axis (with appropriate translation), whereas below the transition this is no longer the case. As a consequence, the two different solutions below the transition generate different scattering profiles for integer wavevectors $(klm)$ with $k\neq l$. Experimental data is available for the $(200)$ peak, and does indeed show two different scattering profiles, labelled according to the direction of field-sweep \cite{Petrenko2002}. It is not clear how the direction of field-sweep should affect the chosen spin configuration, but the profiles found experimentally and those predicted by our model are in good agreement. The predictions of the model could be further tested by looking at the scattering profiles of the $(210)$ and $(201)$ peaks, which also show significant variation between the two solutions in our model. Predicted scattering profiles for various peaks are shown in figure \ref{fig:elasticScatterBz}.

Below $1.6\T$ additional local minima are found in the classical spin configurations. The number of such local minima increases as the field decreases, and hundreds are found for fields less than $0.9\T$. The energy gap between these local minima and the ground state decreases as the field decreases, falling below $1 \K$ at a field of $0.6 \T$ and below $0.1 \K$ at a field of $0.2\T$. Experimental studies show a second phase transition at a field of around $0.7\T$ \cite{Hov1980,Rousseau2017}. While the model presented here does not show such a transition, the increasing number of configurations in a narrow band of energies around the ground state is consistent with the development of a spin-liquid phase. Simulations on larger lattice sizes may be able to show this more clearly. We note that in our model the minimum energy classical solutions remain stable to spinwave excitations at low fields, so the transition at $0.7\T$ cannot be understood in the same way as the transition from the high-field phase, i.e. it does not appear to be driven by a magnon mode becoming soft. This is consistent with experimental evidence of latent heat, suggesting a first order transition \cite{Tsui1999}.

\section{Discussion}\label{sec:discuss}

The treatment of the model presented in this paper gives a description of the transition from the high-field paramagnetic phase of gadolinium gallium garnet to the lower field intermediate phase as a breaking of the BCC-periodicity of the ground-state spins, caused by one or more magnon modes at incommensurate wavevectors becoming soft. The dependence of the critical field and the nature of the intermediate phase on the direction of the magnetic field is shown to be a result of the significant dipole-dipole interactions, which distort the lowest magnon band and thus determine the wavevectors and energies of the minimum energy modes.

Predicted values of the critical field are higher than the experimentally observed values by a factor of approximately $1.25$, and agreement with experimental neutron scattering profiles above and below the transition is improved if the external field in the model is linearly scaled by the same factor. The simplest way to account for this difference between theory and experiment would be to change the antiferromagnetic coupling parameters of the model. A reduction of $J_1$ by around 25\% would align the predicted critical fields with those seen in neutron scattering experiments, and also reduces the energy gap between the lowest and highest magnon bands to more closely align with experimental data (compare figure \ref{fig:inelasticScatteringPowder} of this paper with figures 2 and 3 of \cite{dAmbrumenil2015}). The value of $J_1 = 0.107\K$, used in nearly all theoretical studies to date is based upon \cite{Kinney1979}, which used magnetic susceptibility measurements to impose two non-linear conditions on the parameters $(J_1,J_2,J_3)$. The reported uncertainty in the value of $J_1$ given the values of the other parameters is rather large, and a 25\% reduction of $J_1$ is (just) consistent with the measurements, provided that the value of $J_3$ is increased correspondingly (see figure 3 of \cite{Kinney1979}). Data from subsequent studies which have assumed the value $J_1 = 0.107\K$ could probably be used to reduce the uncertainty in the parameter values significantly. Given the vast increase in experimental data for the gadolinium gallium garnet system since these estimates were first made, there is a case to revisit the estimation of the $J_1$ parameter, for example as was done for $J_2$ and $J_3$ in \cite{Yavorskii2006}.

A previously reported Monte-Carlo simulation of the same system found a similar discrepancy in the predicted and observed critical fields, which was attributed to the effects of quantum fluctuations and disorder in the real physical samples \cite{Schiffer1994}. We can obtain an estimate for the size of quantum effects by comparing the classical $|S|=7/2$ with the ground state expectation value of the $|\langle \hat{S}^z_{A,i} \rangle |$ predicted by our model. Such a calculation shows a reduction of at most 0.5\% above the transition and at most 3\% below the transition (for $B||(001)$). While the magnon-magnon interactions corresponding to higher order terms in the Holstein-Primakoff expansion are not included in the model presented here, it seems unlikely that they would be sufficient to explain the observed difference in critical field. A similar treatment of rare-earth ions in a pyrochlore lattice also concluded that quantum corrections were small, although the dipole-dipole interaction was found to increase the size of the fluctuations \cite{DelMaestro2004}.

As noted above, the inclusion of the dipole-dipole interaction significantly increases the predicted value of the critical field, from approximately $1.7\T$ in the model with only exchange and Zeeman terms, to around $2.0\T$. This is in line with the finding that the dipole interaction generally acts to soften the magnon modes \cite{DelMaestro2004}. The Ewald method used here is correct for a spherical sample in a vacuum, and the geometries and environments of real systems will introduce corrections to the calculation \cite{Wang2001}. For ellipsoidal crystals the internal magnetic field can be written as $\mathbf{B}_i = \mathbf{B} - \mu_0 \mathbf{N}\cdot \mathbf{M}$ where $\mathbf{M}$ is the magnetization and $\mathbf{N}$ is the magnetic depolarization tensor determined by the geometry. For a spherical crystal, $\mathrm{N}=1/3$ (scalar), whereas for needle-like crystals such as those used in \cite{Schiffer1994}, approximation by an infinitely long cylinder gives $\mathbf{N}$ with eigenvalues $\mathrm{N}_\perp = 1/2$ perpendicular to the long axis and $\mathrm{N}_\parallel=0$ parallel to the long axis \cite{Osborn1945}. For a fully polarized crystal we find $\mu_0 |\mathrm{M}| \approx 1.0\T$ so this gives a maximum difference of around $0.3\T$ in the internal magnetic field between these two limiting cases. Further work is necessary to fully understand these effects. 

One of the most intriguing features of the intermediate-field phases is the existence of strong Bragg scattering peaks seen at several incommensurate wavevectors \cite{Petrenko2009}. The model presented here is unable to provide an explanation for the position and intensities of these peaks. To do so may require dealing with the many nearly-degenerate spin configurations that are found to exist on larger lattices at lower fields. On a smaller scale, extending the model to spin configurations with a periodicity of $2 \times 2 \times 2$ unit cubic cells would enable the study of half-integer peaks also seen in experiments \cite{Deen2015,Petrenko2009}.

\section*{Acknowledgements}

M. A. would like to thank Oleg Petrenko for useful discussions. This research was supported by The Catholic University of Korea Research Fund, 2020, and by the Basic Science Research Program through the National Research Foundation of Korea(NRF), Grant Number 2017R1C1B5017323.

\section*{References}

\bibliographystyle{unsrt.bst}
\bibliography{GGG2020_08}

\begin{thebibliography}{10}

\bibitem{Hov1980}
S.~Hov, H.~Bratsberg, and A.~T. Skjeltorp.
\newblock Magnetic phase diagram of gadolinium gallium garnet.
\newblock {\em J. Mag. Mag. Mat.}, 15-18:455, 1980.

\bibitem{Ramirez1991}
A.~P. Ramirez and R.~N. Kleiman.
\newblock Low-temperature specific heat and thermal expansion in the frustrated
  garnet {$\mathrm{Gd}_3 \mathrm{Ga}_5 \mathrm{O}_{12}$}.
\newblock {\em J. Appl. Phys.}, 69:5252, 1991.

\bibitem{Tsui1999}
Y.~K. Tsui, C.~A. Burns, J.~Snyder, and P.~Schiffer.
\newblock Magnetic field induced transitions from spin glass to liquid to long
  range order in a {3D} geometrically frustrated magnet.
\newblock {\em Phys. Rev. Lett.}, 82(17):3532, 1999.

\bibitem{Deen2015}
P.~P. Deen, O.~Florea, E.~Lhotel, and H.~Jacobsen.
\newblock Updating the phase diagram of the archetypal frustrated magnet
  {$\mathrm{Gd}_3 \mathrm{Ga}_5 \mathrm{O}_{12}$}.
\newblock {\em Phys. Rev. B}, 91(1):014419, 2015.

\bibitem{Rousseau2017}
A.~Rousseau, J.-M. Parent, and J.~A. Quilliam.
\newblock Anisotropic phase diagram and spin fluctuations of the hyperkagome
  magnet {$\mathrm{Gd}_3 \mathrm{Ga}_5 \mathrm{O}_{12}$} as revealed by sound
  velocity measurements.
\newblock {\em Phys. Rev. B}, 96(6):060411, 2017.

\bibitem{Paddison2015}
J.~A.~M. Paddison, H.~Jacobsen, O.~A. Petrenko, M.~T. Fern{\'a}ndez-Diaz, P.~P.
  Deen, and A.~L. Goodwin.
\newblock Hidden order in spin-liquid {$\mathrm{Gd}_3 \mathrm{Ga}_5
  \mathrm{O}_{12}$}.
\newblock {\em Science}, 350(6257):179, 2015.

\bibitem{Schiffer1995}
P.~Schiffer, A.~P. Ramirez, D.~A. Huse, P.~L. Gammel, U.~Yaron, D.~J. Bishop,
  and A.J.Valentino.
\newblock Frustration induced spin freezing in a site-ordered magnet:
  Gadolinium gallium garnet.
\newblock {\em Phys. Rev. Lett.}, 74(12):2379, 1995.

\bibitem{Petrenko1998}
O.~A. Petrenko, C.~Ritter, M.~Yethiraj, and D.~McK Paul.
\newblock Investigation of the low-temperature spin-liquid behavior of the
  frustrated magnet gadolinium gallium garnet.
\newblock {\em Phys. Rev. Lett.}, 80(20):4570, 1998.

\bibitem{Quilliam2013}
J.~A. Quilliam, S.~Meng, H.~A. Craig, L.~R. Corruccini, G.~Balakrishnan, O.~A.
  Petrenko, A.~Gomez, S.~W. Kycia, M.~J.~P. Gingras, and J.~B. Kycia.
\newblock Juxtaposition of spin freezing and long range order in a series of
  geometrically frustrated antiferromagnetic gadolinium garnets.
\newblock {\em Phys. Rev. B}, 87(17):174421, 2013.

\bibitem{Ghosh2008}
S.~Ghosh, T.~F. Rosenbaum, and G.~Aeppli.
\newblock Macroscopic signature of protected spins in a dense frustrated
  magnet.
\newblock {\em Phys. Rev. Lett.}, 101(15):157205, 2008.

\bibitem{dAmbrumenil2015}
N.~d'Ambrumenil, O.~A. Petrenko, H.~Mutka, and P.~P. Deen.
\newblock Dispersionless spin waves and underlying field-induced magnetic order
  in gadolinium gallium garnet.
\newblock {\em Phys. Rev. Lett.}, 114(22):227203, 2015.

\bibitem{Schiffer1994}
P.~Schiffer, A.~P. Ramirez, D.~A. Huse, and A.~J. Valentino.
\newblock Investigation of the field induced antiferromagnetic phase transition
  in the frustrated magnet: Gadolinium gallium garnet.
\newblock {\em Phys. Rev. Lett.}, 73(18):2500, 1994.

\bibitem{Petrenko2002}
O.~A. Petrenko, G.~Balakrishnan, D.~McK Paul, M.~Yethiraj, and J.~Klenke.
\newblock Field-induced transitions in the highly frustrated magnet gadolinium
  gallium garnet –- long- or short-range order?
\newblock {\em Appl. Phys. A}, 74(Suppl.):S760, 2002.

\bibitem{Petrenko1999}
O.~A. Petrenko, D.~McK Paul, C.~Ritter, T.~Zeiske, and M.~Yethiraj.
\newblock Magnetic frustration and order in gadolinium gallium garnet.
\newblock {\em Physica B}, 266:41, 1999.

\bibitem{Petrenko2009}
O.~A. Petrenko, G.~Balakrishnan, D.~McK Paul, M.~Yethiraj, G.~J. McIntyre, and
  A.~S. Wills.
\newblock Field induced magnetic order in the frustrated magnet gadolinium
  gallium garnet.
\newblock {\em J. Phys.: Conf. Ser.}, 145:012026, 2009.

\bibitem{Kinney1979}
W.~I. Kinney and W.~P. Wolf.
\newblock Magnetic interactions and short range order in gadolinium gallium
  garnet.
\newblock {\em J. Appl. Phys.}, 50(3):2115, 1979.

\bibitem{Overmeyer1963}
J.~Overmeyer.
\newblock {\em Paramagnetic Resonance}, chapter~15.
\newblock Academic Press, New York, 1963.

\bibitem{Yavorskii2006}
T.~Yavors{'}kii, M.~Enjalran, and M.~J.~P. Gingras.
\newblock Spin hamiltonian, competing small energy scales, and incommensurate
  long-range order in the highly frustrated {$\mathrm{Gd}_3 \mathrm{Ga}_5
  \mathrm{O}_{12}$} garnet antiferromagnet.
\newblock {\em Phys. Rev. Lett.}, 97(26):267203, 2006.

\bibitem{Petrenko2000}
O.~A. Petrenko and D.~McK Paul.
\newblock Classical heisenberg antiferromagnet on a garnet lattice: A monte
  carlo simulation.
\newblock {\em Phys. Rev. B}, 63(2):024409, 2000.

\bibitem{Yavorskii2007}
T.~Yavors{'}kii, M.~J.~P. Gingras, and M.~Enjalran.
\newblock Ill-behaved convergence of a model of the {$\mathrm{Gd}_3
  \mathrm{Ga}_5 \mathrm{O}_{12}$} garnet antiferromagnet with truncated
  magnetic dipole–dipole interactions.
\newblock {\em J. Phys. Condens. Matter}, 19:145274, 2007.

\bibitem{Ewald1921}
P.~P. Ewald.
\newblock Die berechnung optischer und elektrostatischer gitterpotentiale.
\newblock {\em Annalen der Physik}, 369(3):253, 1921.

\bibitem{deLeeuw1980}
S.~W. de~Leeuw, J.~W. Perram, and E.~R. Smith.
\newblock Simulation of electrostatic systems in periodic boundary conditions.
  {I. Lattice sums and dielectric constants}.
\newblock {\em Proc. R. Soc. A}, 373:27, 1980.

\bibitem{Wang2001}
Z.~Wang and C.~Holm.
\newblock Estimate of the cutoff errors in the {E}wald summation for dipolar
  systems.
\newblock {\em J. Chem. Phys.}, 115(14):6351, 2001.

\bibitem{Bergman2008}
D.~L. Bergman, C.~Wu, and L.~Balents.
\newblock Band touching from real-space topology in frustrated hopping models.
\newblock {\em Phys. Rev. B}, 78(12):125104, 2008.

\bibitem{DelMaestro2004}
A.~G.~Del Maestro and M.~J.~P. Gingras.
\newblock Quantum spin fluctuations in the dipolar heisenberg-like rare earth
  pyrochlores.
\newblock {\em J. Phys.: Condens. Matter}, 16:3339, 2004.

\bibitem{Gingras2004}
M.~Enjalran and M.~J.~P. Gingras.
\newblock Theory of paramagnetic scattering in highly frustrated magnets with
  long-range dipole-dipole interactions: The case of the
  {$\mathrm{Tb}_2\mathrm{Ti}_2\mathrm{O}_7$} pyrochlore antiferromagnet.
\newblock {\em Phys. Rev. B}, 70(17):174426, 2004.

\bibitem{Jensen1991Book}
J.~Jensen and A.~R. Mackintosh.
\newblock {\em Rare Earth Magnetism: Structures and Excitations}.
\newblock Clarendon Press, Oxford, 1991.

\bibitem{Osborn1945}
J.~A. Osborn.
\newblock Demagnetizing factors of the general ellipsoid.
\newblock {\em Physical Review}, 67(11-12):3351, 1945.

\end{thebibliography}

\end{document}